\documentclass[twoside]{article}

\usepackage{url,intmacros,graphicx}
\usepackage{amssymb,amsmath}
\usepackage{capt-of}
\usepackage{float}

\def\BibTeX{{\rm B\kern-.05em{\sc i\kern-.025em b}\kern-.08em
    T\kern-.1667em\lower.7ex\hbox{E}\kern-.125emX}}

\numberwithin{equation}{section}

\newcommand{\email}[1]{\\ \small{\url{#1}} \\}
\newcommand{\institution}[1]{\\ \parbox{3.0in}{\small{#1}}}
\newcommand{\keywords}[1]{\small\textbf{Keywords: }#1}
\newcommand{\AMSsubj}[1]{\noindent\small\textbf{AMS subject classifications: }#1}
\newcommand{\CCSconcepts}[1]{\noindent\small\textbf{ACM CCS concepts: }#1}
\newcommand\whenaccepted{}

\title{Concurrent Processing Memory \footnote{\whenaccepted}}

\author{Chengpu Wang
\institution{40 Grossman Street, Melville, NY 11747, USA}
\email{Chengpu@gmail.com}}

\begin{document}
\maketitle
\begin{abstract}
A theoretical memory that embeds limited, application-specific processing power and nearest-neighbor connectivity at every storage element is proposed.  
Such a memory performs parallel computation within itself to solve generic array problems, while remaining pin- and function-compatible with conventional random-access memory.  
The applicability of this in-memory, finest-grain, massive-SIMD approach is examined in detail through a family of increasingly capable devices---content movable, searchable, value-comparable, and computable memory.  
For an array of $N$ items, the approach reduces the instruction-cycle count of universal operations (insertion, deletion, and match finding) to $\sim 1$, of local operations (filtering and template matching) to $\sim$ the operation footprint, and of global operations (summation and finding minimum/maximum) to $\sim \sqrt{N}$.  
It eliminates most data-processing traffic on the system bus, yet remains general-purpose, easy to program, backward-compatible with existing bus-sharing architectures and operating systems, and practical to implement along a clear road map.
\end{abstract}

\keywords{processing-in-memory; in-memory computing; massive SIMD; content-addressable memory; associative processing; smart memory; near-data processing; array processing; memory architecture; von~Neumann bottleneck}

\CCSconcepts{$\bullet$~Computer systems organization $\rightarrow$ Parallel architectures $\rightarrow$ Single instruction, multiple data;\ $\bullet$~Hardware $\rightarrow$ Emerging technologies $\rightarrow$ Emerging architectures;\ $\bullet$~Hardware $\rightarrow$ Memory and dense storage.}

\AMSsubj{68M07 (computer architecture); 68W10 (parallel algorithms); 68P05 (data structures); 68M20 (performance evaluation of systems).}

\section{Introduction}

The array is the most pervasive data structure in computer engineering. 
\begin{itemize}
\item The homogeneous array is the most prevalent collection in most computer languages~\cite{dershem1990}.

\item Text is an array of characters~\cite{stringsearch}.

\item A relational-database table is itself an array of rows of identical column structure~\cite{ramakrishnan2002}.

\item Images and video are arrays of pixels~\cite{davies1990,offen1986}.

\item Scientific and engineering models discretize continuous fields onto arrays of cells or nodes~\cite{szabo1991}.

\item Tensors that dominate modern machine learning are multidimensional arrays~\cite{lecun2015}.
\end{itemize}
Consequently, a large fraction of real computation consists of applying identical, repeated operations to the identical, repeated elements of an array---insertion and deletion, filtering and template matching, summation, limit finding, sorting, and search.  
Such array operations are inherently data-parallel: logically, the same operation applies to every element at once, which makes them the natural target of any parallel architecture.

The sustained growth of transistor budgets under Moore's law~\cite{moore1965} has reopened a long-running debate over whether to invest them in instruction-level~\cite{patt1997,burger2004}, thread-level~\cite{hammond2000}, or data-level parallelism~\cite{soliman2010}.  
Data-level parallelism is the simplest of the three in both programmability and hardware construct, and the single-instruction-multiple-data (SIMD) model remains its natural expression.  
Over the past two decades, SIMD has moved from the periphery to the mainstream of high-performance computing.  
The graphics processing unit (GPU) has become the dominant data-parallel engine: its single-instruction-multiple-thread (SIMT) execution model applies one instruction stream across thousands of lightweight cores~\cite{lindholm2008,nickolls2010}, and it now underpins scientific computing, image and signal processing~\cite{owens2008}, and the training and inference of large-scale machine-learning models~\cite{lecun2015}.  
Vector extensions inside conventional CPUs (SSE, AVX, and their successors)~\cite{sse} bring the same principle, at a smaller scale, into general-purpose processors.

Most large-scale parallelism, however, is organized as multiple-instruction-multiple-data (MIMD) execution---shared-memory multiprocessors and cluster or grid systems~\cite{magoules}---standardized in software as OpenMP~\cite{openmp} and MPI~\cite{mpi}, with a large literature devoted to their scalability and performance tuning for scientific computing~\cite{heroux2006,labarta2006, oliker2006, almasi2006, teresco2006}.  
Yet mapping an inherently data-parallel problem onto an MIMD fabric incurs communication and synchronization overheads that a true SIMD execution avoids.

Both of these mainstream directions, however, keep computation and storage physically separate, so that data must still stream across an interconnect for processing~\cite{hayes1988,hennessy1998,culler1999}.  
This separation is the enduring bottleneck of the von Neumann architecture: processor throughput has for decades grown far faster than memory bandwidth and latency---the ``memory wall''~\cite{wulf1995}---and the movement of data, rather than the arithmetic performed on it, now dominates both the execution time and the energy budget of data-intensive workloads, with an off-chip memory access costing orders of magnitude more energy than the operation it feeds~\cite{horowitz2014}.  
For the past 20 years, the CPU/GPU clock speed has stalled at a few GHz~\cite{horowitz2014}, while the speed of fetching from main memory has barely changed, at about 0.3~GHz~\cite{wulf1995}.
To cope with this ten-fold difference, first- and second-level caches are used to preload and buffer instructions and data~\cite{hennessy1998}, at the cost of a vast amount of wasted execution, such as after a branch misprediction, when all but one speculative path are abandoned.
The prevailing MIMD approaches only aggravate this bus-bottleneck problem~\cite{culler1999}.

A very different paradigm of massive parallelism is emerging in quantum computing~\cite{nielsen2010}, which exploits superposition and entanglement to evaluate a function over an exponentially large input space in a single coherent step~\cite{shor1997}.  
While a quantum computer is not a SIMD machine---its parallelism is probabilistic and must be extracted through interference and measurement rather than read out directly---it shares with SIMD the ambition of amortizing a single control action across a vast amount of data, and it illustrates the continuing search for architectures that break the serial, bus-bound execution of conventional computers.
Although quantum computing is markedly superior for niche applications such as decryption~\cite{shor1997}, it is not expected to replace traditional von Neumann machines for the majority of applications.

This work~\cite{wang2003} instead pursues data-level parallelism inside the memory itself, so that array operations execute where the data already reside.
The wider field has moved in the same direction.  
Beyond simply widening the interface with high-bandwidth stacked memory, processing-in-memory has advanced from proposal to product---DRAM augmented with programmable compute units~\cite{kwon2021}, general-purpose in-memory processor cores~\cite{devaux2019}, and a broad body of near- and in-memory architectures~\cite{mutlu2022}.  
These systems confirm the premise of this work, that the bus bottleneck calls for in-memory processing; yet they have largely adopted coarse-grain near-memory processing or analog compute rather than the fine-grain, RAM-compatible, per-element SIMD explored here.  

The proposed CPM (Concurrent Processing Memory) therefore occupies a still-unexplored point in a design space whose importance the field has since confirmed.
It incorporates many features of existing SIMD, associative, and in-memory architectures such as CAM~\cite{chivin1989}, associative processors~\cite{foster1976, krikelis1997, batcher1974, weems1989, lea1988, rangan2001, walker2001}, and PIM~\cite{diva1999, flexram2003, iram, smartmemories2000, oskin2000, tennenhouse1997}, and provides three unique improvements:
\begin{itemize}
\item It is pin- and function-compatible with a traditional RAM in a conventional von Neumann architecture.

\item It uses an array decoder to allow an arbitrary size for each array item.

\item It defines a family of CPM as a road map for adoption.
\end{itemize}
CPM achieves SIMD with the finest-grain processor without data movement, in contrast to the prevailing GPU, which is the most successful SIMD engine to date.
A GPU achieves its throughput by moving data into a dedicated device memory, processing it with thousands of SIMT cores, and moving the results back~\cite{lindholm2008,nickolls2010}; it therefore hides, rather than removes, the cost of data movement, and its efficiency falls sharply on irregular or low-arithmetic-intensity workloads.
The two approaches are complementary---a GPU excels at dense numerical kernels, whereas CPM targets the universal, local, and global array operations that dominate database, search, and image-processing preprocessing workloads.

\section{CPM (Concurrent Processing Memory)}

\subsection{Architecture}

\begin{figure}[htbp]
\centering
\includegraphics[width=0.8\linewidth]{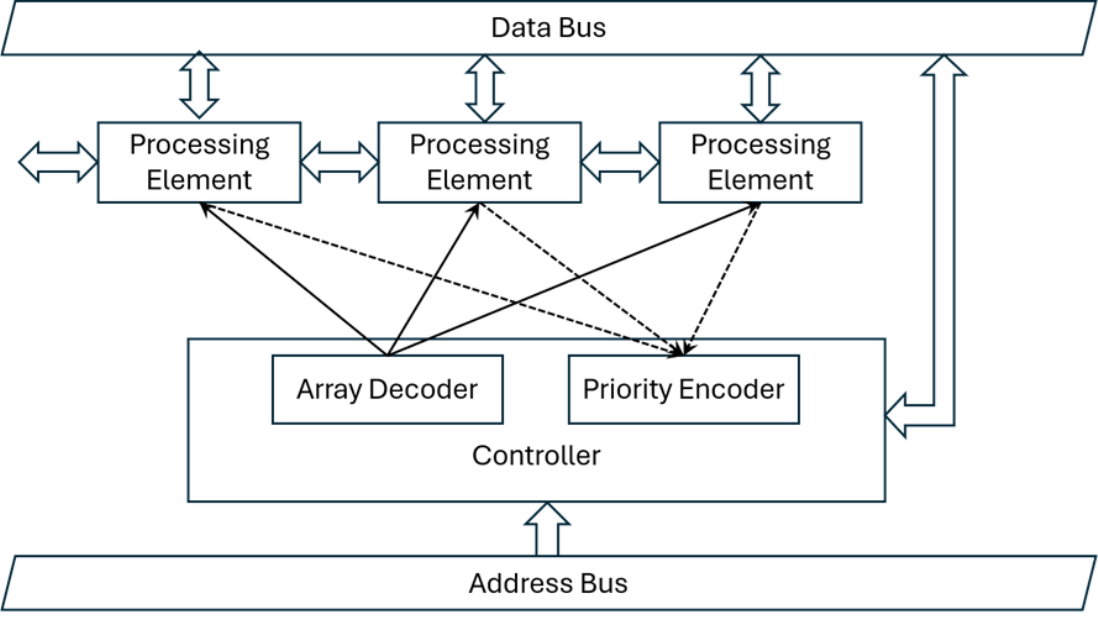}
\caption{Concurrent Processing Memory architecture.}
\label{fig:cpm}
\end{figure}

CPM is designed to work with a bus~\cite{hayes1988} in a conventional von Neumann architecture, which consists of an address bus and a data bus.
As shown in Figure~\ref{fig:cpm}, the basic rules for CPM are:
\begin{enumerate}
\item\label{rule:identical} A CPM is made of identical PEs (Processing Elements), each of which contains a fixed number of registers.

\item\label{rule:address} Each PE has an addressable register, allowing read and write operations to the addressable register through the data bus.

\item\label{rule:connect} Neighboring PEs are connected so that a PE can read a neighboring register of each of its neighbors.
The content of the addressable register and the neighboring register can be copied in either way.

\item\label{rule:controller} In addition to PEs, a CPM also contains a controller.
Some of the registers in the controller are also accessible by the address bus and the data bus.

\item\label{rule:activate} Multiple PEs can be activated concurrently by an array decoder inside the controller if each of their element addresses is (1) no less than a start address, (2) no more than an end address, and (3) an integer increment starting from the start address.

\item\label{rule:identify} Multiple activated PEs can identify themselves concurrently to a priority encoder~\cite{hayes1988} inside the controller.

\item\label{rule:execute} Multiple activated PEs can read and execute the same instruction concurrently from the address bus and the data bus.
\end{enumerate}

All array items form a periodic pattern when they are saved continuously in a memory.  
Rule~\ref{rule:activate} allows the storing of each array item by a PE or multiple neighboring PEs.

Except for Rule~\ref{rule:activate}, all of the above rules have mature usages:
\begin{itemize}
\item Rules~\ref{rule:identical} and~\ref{rule:execute} achieve SIMD concurrency \cite{fountain1994}.  

\item Rule~\ref{rule:address} specifies the functional backward compatibility with a conventional RAM~\cite{hayes1988}.  

\item Rule~\ref{rule:connect} provides the nearest-neighbor connectivity found in systolic and mesh-connected processor arrays \cite{kung1982,culler1999}.

\item Rule~\ref{rule:controller} is a common technique in programming devices via the address bus and data bus in a conventional von Neumann architecture, such as using an extra address bit to indicate whether the data bus contains a datum or an instruction to each PE \cite{hayes1988}.

\item Rule~\ref{rule:identify} is used extensively in a content-addressable memory \cite{chivin1989}.
\end{itemize}

\subsection{PE Capability}

As a massive-SIMD approach, the silicon budget \cite{rabaey1999} of each PE needs to be controlled carefully.  
CPM is actually a family name comprising members in the order of PE complexity: 
\begin{enumerate}
\item CMM (Content Movable Memory): a memory with object packing capability, and with static RAM (SRAM) performance but dynamic RAM (DRAM) construct~\cite{rabaey1999,hennessy1998}.

\item CSM (Content Searchable Memory): In addition to being a CMM, it is a super CAM (Content Addressable Memory) with arbitrary operand widths.

\item CVM (Content Value-comparable Memory): A CSM with value comparison capability.

\item CCM (Content Computable Memory): A CVM with arithmetic capability for arbitrary operand widths.
\end{enumerate}
Except for CMM, each type of CPM targets a classic and ubiquitous array problem only.
It is possible that a bus-sharing system contains multiple types of CPM, such as CMM for general memory management, CSM for match search, CVM for value-comparison search, and CCM for simple but massive value calculations, which may all hold different parts of one record.
The CPM of simpler PE construct is also simpler to implement, and the complexity progression from CMM to CCM can be used as a roadmap to introduce CPM into a modern bus-sharing system.

\subsection{Topology}

A silicon device is usually constructed on a square area~\cite{rabaey1999}.
It is reasonable to assume that a CPM of $N$ PEs is constructed as a square array of $\sqrt{N} \times \sqrt{N}$ PEs, with each PE having four nearest neighboring PEs, except those at the boundary.
The prevailing layered fabrication techniques allow a multi-layered construct~\cite{threedic}, giving each internal PE six nearest neighboring PEs and the corresponding connections.
The connection between two PEs on the opposite boundaries is generally more expensive in both construct and performance.
In contrast, a zig-zag packaging guarantees one-dimensional neighboring connectivity at the expense of strange two- or three-dimensional neighboring connectivity.
Also, the controller can be a CPU with a super-wide bus, supporting activation far more complicated than Rule~\ref{rule:activate}.
Therefore, Rules~\ref{rule:connect} and~\ref{rule:activate} are huge simplifications for the theoretical discussion in this study.

\section{CMM (Content Movable Memory)}

A CMM is the simplest CPM:
\begin{itemize}
\item Rule~\ref{rule:activate} is simplified to have only an address increment of 1.

\item Rule~\ref{rule:identify} is not implemented.
\end{itemize}

Using Rule~\ref{rule:connect} and the simplified Rule~\ref{rule:activate}:
\begin{itemize}
\item The content of the CMM can be moved by concurrently copying the content of the addressable register to the addressable register of neighboring PEs through the neighboring register.

\item By moving the content concurrently, the object size in a CMM can be changed dynamically, so that a CMM has the most efficient memory usage with neither a pre-allocation requirement nor the associated memory overflow, providing much simplified memory management~\cite{hayes1988, hennessy1998}.

\item The addressable registers and the neighboring registers can be made of DRAM cells.
The content of the addressable registers can be updated by first copied to the neighboring register then copied back.
Because a neighboring register bit only needs to hold it content for very short duration, its capacitance can be much smaller and its silicon area is negligible.
With the content of each addressable register readily available, a CMM can have the performance of a SRAM, but the cost of a DRAM.
\end{itemize}
CMM is aimed at replacing both SRAM and DRAM due to its memory-management functionality and its overall simplicity, speed, and density.
It may fundamentally change the programming and software industry.

One additional benefit of CMM is that it can be made fault-tolerant~\cite{siewiorek1998} when each neighboring register holds a copy of its addressable register's content.

\section{CSM (Content Searchable Memory)}

A CSM is similar to a CAM (Content Addressable Memory)~\cite{chivin1989,pagiamtzis2006}: it matches and identifies all activated addressable registers against the content on the data bus.
Each PE has one additional status bit to save the result of the previous match, which can be read by neighboring PEs for more complicated matches.
In this way, the CSM removes the length and alignment limits of CAM.
A search involving $M$ neighboring PEs takes $\sim M$ instruction cycles.

The priority encoder enables enumeration of all matches.
In addition to the priority encoder, the controller of a CSM may contain other functional units, such as a parallel counter to count the total number of matches.
Like a CAM, a CSM reduces a match search from $\sim N$ operations to $\sim 1$ operation.

In a conventional bus-sharing architecture, resolving which records hold a given value requires an $\sim N$ scan, which is why the values to be frequently searched need to be indexed beforehand.
Indexing underlies not only value lookups, but also the joins and foreign-key relationships that define relations among tables in a relational database~\cite{ramakrishnan2002, stephens2002}.  
Every such index must be stored, kept consistent on every insertion, deletion, and update, and is often dropped before a bulk update and rebuilt afterward.  
A CSM eliminates indexing altogether: value-to-location resolution---and hence lookups, joins, and foreign-key reference resolution---is performed by the memory itself in $\sim 1$ operation \emph{in situ}, with no pre-processing, no auxiliary index table, and no maintenance overhead.

\section{CVM (Content Value-comparable Memory)}

A CVM extends the functionality of a CSM from value matching to value comparing.

\subsection{Find Threshold Values}

Finding all addressable values below or above one threshold value on the data bus needs only $\sim 1$ instruction cycle.

\subsection{Find Local Minimum or Local Maximum}

By comparing each value with its two neighbors, all values that are either a local minimum or a local maximum are found in $\sim 1$ instruction cycle.

\subsection{Find Global Minimum or Global Maximum}

\begin{figure}[htbp]
\centering
\includegraphics[width=0.6\linewidth]{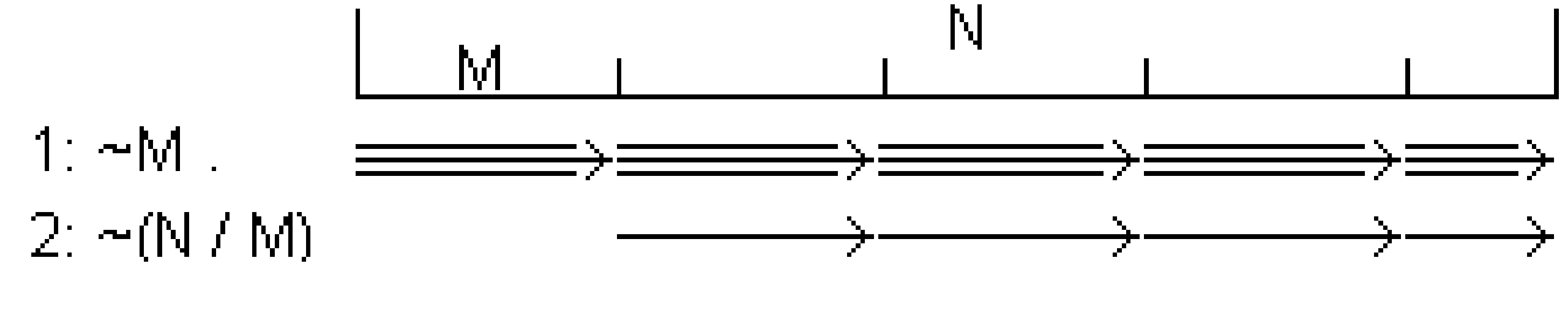}
\caption{The execution flow diagram of a global operation on a 1-D CPM with $N$ array items.}
\label{fig:global}
\end{figure}

Finding the global minimum or global maximum for $N$ array items takes $\sim \sqrt{N}$ instruction cycles.
Let the increment count be $M$. In figure \ref{fig:global}:
\begin{enumerate}
\item From the starting address plus one, for all the $N/M$ array items concurrently, copy the larger or smaller content of the addressable register and the addressable register of the left PE to the neighboring registers.

\item Repeat the above step $M$ times, to find the local minimum or maximum within each of the $N/M$ sections.

\item Copy the content of the neighboring registers to the corresponding addressable registers.

\item Poll the $N/M$ section minimums or maximums serially over the address bus and data bus to obtain the global minimum or maximum.

\item The total instruction cycles are $\sim M + N/M$, which has a minimum of $\sim \sqrt{N}$ when $M = \sqrt{N}$.
\end{enumerate}

\subsection{Sorting Array}

Inserting a new array item into an already sorted array takes $\sim 1$ instruction cycle, so loading an array to a CVM will result in a sorted array.

Moving a value takes $\sim 1$ instruction cycle, and the position of the array item is relatively correct after one such move.
Detecting if an array is already sorted takes $\sim 1$ instruction cycle.
Sorting an array takes $\sim N$ instruction cycles in the worst case, and $\sqrt{N}$ instruction cycles in the average case.

\subsection{Counting for Statistics}

By enriching the controller with more capability, such as parallel counters, memories, and an ALU (Arithmetic and Logic Unit), the statistical properties of the array such as a histogram can be computed efficiently using Rule~\ref{rule:identify}.

\section{CCM (Content Computable Memory)}

If each PE of a CVM is further enriched with an ALU, the result is a CCM.

\subsection{Local Operations}

Let a special 1-D vector of an odd number of items describe the content of the neighboring register.
It is named as the neighboring vector.
For example:
\begin{itemize}
\item $(1)$ means the content of the addressable register of the PE itself.

\item $(1,0,0)$ means the content of the addressable register of the PE to its left, which is achieved by a left move.

\item $(1,2,1)$ means the Gaussian averaging of the nearest neighbors.

\item $(1,2,4,2,1)$ means a better Gaussian averaging.
\end{itemize}
To achieve $(1,2,1)$:
\begin{enumerate}
\item Add the content of the left addressable register to that of this addressable register, storing the result in the neighboring register: $(1) + (1,0,0) = (1,1,0)$.

\item Add the content of the right neighboring register to that of this neighboring register: $(1,1,0) + (0,1,1) = (1,2,1)$.
\end{enumerate}

For a CCM with multi-dimensional connectivity, the neighboring vector can be extended to a multi-dimensional array or tensor.

The neighboring vector can contain a template, to be matched directly against the expected value on the data bus.

A local operation involving $M$ neighboring PEs takes $\sim M$ instruction cycles.

\subsection{Encryption and Decryption}

The combined content of neighboring addressable registers can be reversed if the neighboring vector is known, but the vector is difficult to infer from the result alone.
Therefore, it can be used for encryption and decryption.

\subsection{Sum}

The algorithm of finding global minimum or global maximum can be modified to sum an array in $\sim \sqrt{N}$ instruction cycles.

To sum a two-dimensional array of $N_x \times N_y$ items, the minimal total instruction cycle count is $\sim \sqrt[3]{N_x N_y}$ when $M_x \sim M_y \sim \sqrt[3]{N_x N_y}$.

\subsection{Context Switch}

To justify the cost of the ALU, each PE may contain multiple addressable registers, and allow some of the registers to load the contexts of other jobs.

\subsection{Edge Detection}

\begin{figure}[htbp]
\centering
\includegraphics[width=0.4\linewidth]{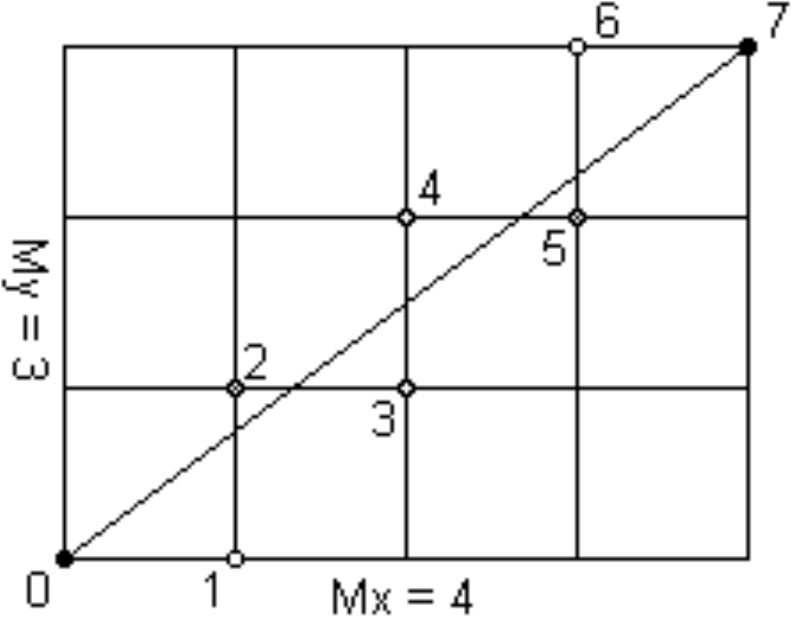}
\caption{2-D line detection using a messenger.}
\label{fig:messenger}
\end{figure}

Due to neighbor-to-neighbor connectivity, a 2-D CCM can treat the edge detection problem~\cite{davies1990} as a neighbor-counting problem.
To detect edge lines with pixel length $L$ lying exactly along the $X$ direction to the left of each pixel, the neighbor-count algorithm is direct:
\begin{enumerate}
\item All pixels concurrently subtract the value of the addressable registers of their bottom PEs from that of their top PEs, and store the result in the neighboring registers.

\item All pixels concurrently sum the neighboring registers of their $L$ left neighbors together with their own.  
The absolute value of the result indicates the possibility of an edge line starting from that pixel, while the sign of the result indicates whether the edge is rising or falling along the $Y$ direction.
\end{enumerate}

To detect edge lines with a slope of $(M_y/M_x)$, in which $M_x$ and $M_y$ are two integers, each pixel defines a rectangular $M_x \times M_y$ pixel area, and the line which connects the pixel and the furthest corner of the area has the slope of $(M_y/M_x)$.  
Similar to obtaining the section sums in a sum algorithm, a messenger starts from the furthest corner of the area and walks $(M_x + M_y)$ steps along the line until it reaches the original pixel.  
At each of its stops, in a predefined fashion, if the pixel is on the left side of the line its intensity is added to the messenger; otherwise its intensity is subtracted from the messenger. 
When reaching the original pixel, the value of the messenger indicates the possibility and the slope direction of the edge-line segment which connects the original pixel and the furthest corner of the $M_x \times M_y$ area.  
Thus, it is called the line-segment value of the pixel for the $M_x \times M_y$ area.  
This accumulating process is carried out concurrently for all the pixels of the image, independent of image size.  
Figure~\ref{fig:messenger} shows the $(4 \times 3)$ area used to detect a line with a slope of $(3/4)$ passing the original pixel at~0.  
The accumulation processing is from pixel~7 to pixel~0 in sequence, with the raw intensity of pixels~1,~3, and~5 added to, and those of pixels~2,~4, and~6 subtracted from, the messenger.

\begin{figure}[htbp]
\centering
\includegraphics[width=0.4\linewidth]{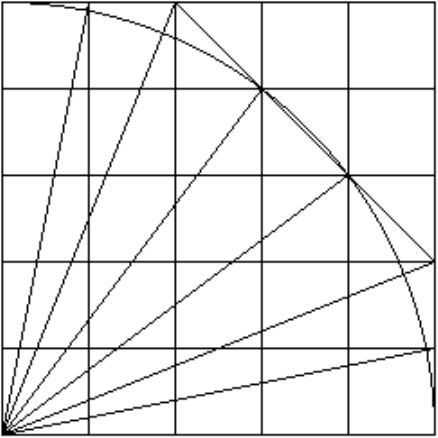}
\caption{A set of lines with pixel spans of about 5 in real distance.}
\label{fig:lines}
\end{figure}

Given an angular-resolution requirement, a $\{(M_x, M_y)\}$ set can be constructed to detect lines of all slopes on an image.  
To construct such a set for an angular resolution $\sim 2/D$, a circle of radius $D$ is drawn on a square net of pixels, and the pixels in the vicinity of the circle are the starting pixels for the messengers in the $\{(M_x, M_y)\}$ set. 
Figure~\ref{fig:lines} shows such a set of lines with $D$ equal to~5.  
The total instruction count to detect all lines in the set is $\sim D^2$, independent of the image size.  
As a result of line detection, each pixel is marked by the best line-segment value together with its corresponding $M_x \times M_y$ area.

\section{Conclusion and Discussion}

As an in-memory massive-SIMD approach, CPM seems to be able to vastly improve the solutions to typical array problems in the framework of a traditional bus-sharing architecture running a prevailing multitask operating system.  
Each type of CPM is specialized to a particular application in both its hardware construct and software instruction set, while the whole set provides a clear adoption path.
The limited capability of CPM suggests that it can perform the coarse, bulk pre-processing of array data, while a conventional CPU or GPU handles the remaining fine-grained or irregular computation.

\subsection{Software Challenges}

Using CPM implies a fundamental and extensive software rewrite, because it logically changes the von Neumann architecture~\cite{hayes1988}.
CMM greatly simplifies traditional memory-management systems~\cite{silberschatz2001}, with speed closer to that of the CPU, reducing the need for the caching systems, which were originally designed to hide the memory--CPU speed disparity~\cite{hennessy1998}.
Although CSM, CVM, and CCM have quite limited capabilities compared with those of a CPU/GPU, they provide massive pre-processing that renders much traditional code optimization~\cite{dershem1990} unnecessary, while demanding corresponding new optimizations.
The challenge is to find enough practical drivers to justify this large change.

\subsection{Hardware Challenges}

One important drawback of massive SIMD is electric grounding: when a massive number of PEs change state at the same time, they draw a transient current from the power supply and dump a return current into ground, stressing the power-distribution network~\cite{popovich2008}.

The transition voltage $\Delta V$ of discharging of an individual DRAM cell is negligible, because $\Delta V = R\,C\,\frac{d v(t)}{d t} = 6.8 \times 10^{-4}\,\text{V}$, in which:
\begin{itemize}
\item $R = 0.68\,\Omega$ is the resistance of a $100 nm$ long copper wire with a $50 nm \times 50 nm$ cross section~\cite{rabaey1999},

\item $C = 10^{-14}$ F is the typical capacitance of a DRAM cell~\cite{rabaey1999},

\item $v_0 = 1$ V is the typical power-supply voltage~\cite{rabaey1999},

\item $v(t) = v_0 e^{-\frac{t}{\tau}}$ is the discharge voltage, in which $\tau = 10^{-11}$ second is the settle time of a single logic transition at a 10 GHz operational frequency.

\item The equivalent peak current is $I = C\,v_0 / \tau = 10^{-3}$ A.

\item The intrinsic discharge time is sufficiently short: $R\,C = 6.8 \times 10^{-15}\,\text{second} \ll \tau = 10^{-11}\,\text{second}$.
\end{itemize}

The voltage drop due to wire resistance to implement Rule \ref{rule:activate} is estimated as:
\begin{enumerate}
\item The capacitance of a gate is about $0.1$ fF~\cite{rabaey1999}, which is 100-fold smaller than that of a DRAM cell.

\item When the transition voltage $\Delta V$ is limited to less than $0.1$ V, the resistance $R$ of a copper wire is limited to less than $10^{4} \Omega$, which corresponds to a length of only $1.5$ mm if the copper wire has a cross section of $50 nm \times 50 nm$.

\item The intrinsic discharge time of the wire is $\sim 10^{-12}$ seconds, which is quick enough for the speed of $10$ GHz.

\item The corresponding chip size is $1.5 \times \sqrt{2} = 2.1$ mm, which is not large at all.

\item Because the area of a DRAM cell is about $6 (50 nm)^2 = 1.5 \times 10^{-14}\,m^2$~\cite{rabaey1999}, such a CMM contains $\sim 3 \times 10^8$ bits with a speed of 10 GHz.

\item If the activation wires for Rule \ref{rule:activate} are treated as a continuous copper layer, the voltage difference between the corner and the center is $\sim 1.5 \times 10^3$ V.
\end{enumerate}
In reality, the large fan-out for Rule \ref{rule:activate} uses amplifiers (repeaters/buffers)~\cite{rabaey1999} in addition to wires, so that the voltage drops should be much less than the above values.
Still very large fan-out over very long distance is a technical challenge when implementing Rule \ref{rule:activate}.
These limits are not unique to CPM: a modern chip with billions of gates faces the same fan-out, power-delivery, and ground-bounce problems, and solves them with buffered clock trees, area-distributed vertical power delivery, and on-chip decoupling capacitance~\cite{popovich2008}.

Each switching bit draws about $10^{-3}$ W peak power~\cite{chandrakasan1992}, limiting the number of simultaneously switching bits of a CMM to $10^3$ given a 1 W power budget.
The JEDEC refresh interval is about $3 \times 10^{-2}$ seconds~\cite{jedec2012}, while each refresh of $10^3$ bits takes about $1 \times 10^{-10}$ seconds, which gives a total of $3 \times 10^{11}$ bits per chip on a 1 W power budget.
This is a worst case estimate, because in reality, charging and discharging of bits cancel each other in the power layer.

In short, a 1 GB CMM with 10 GHz speed, 1 W power consumption, and 11 mm in size is widely applicable.
When adding new functionality to a CMM to achieve CSM, CVM, and CCM, 50 nm technology is a comfortable starting point for possibly finer technologies.
Although CPM needs completely new design and implementation in hardware, the current technology seems completely feasible.

\section*{Acknowledgements}

As an independent researcher, the author feels indebted for the encouragement and valuable discussions of Dr.\ Nick Tredennick from the \emph{Gilder Technology Report}, Prof.\ Pao-Kuang Kuo from Wayne State University, Prof.\ Sangjin Hong and Prof.\ Tzi-cker Chiueh from SUNY at Stony Brook, and the organizers of PDPTA 2003, Prof.\ Hamid R.\ Arabnia from the University of Georgia in particular.

\end{document}